\documentclass[aps, prl, reprint, 
amsmath, amssymb, showkeys,]{revtex4-2}

\usepackage{graphicx}
\usepackage{dcolumn}
\usepackage{bm}
\usepackage[mathlines]{lineno}
\usepackage{textcomp}
\usepackage{xcolor}

\begin{document}
\preprint{APS/123-PRL}
\title{Long baseline optical interferometric imaging with active phase stabilization}

\author{Joshua J. Collier$^1$}
\author{Leuca Patmore$^1$}
\author{John S. Wallis$^1$}
\author{Elrina Hartman$^1$}
\author{Benjamin P. Dix-Matthews$^{1}$}
\author{David R. Gozzard$^{1,2}$}
 \email{david.gozzard@uwa.edu.au}
\affiliation{$^1$International Centre for Radio Astronomy Research, The University of Western Australia, Crawley WA 6009, Australia}
\affiliation{$^2$School of Physics, Mathematics, and Computing, The University of Western Australia, Crawley WA 6009, Australia}

\date{\today}

\begin{abstract} 
Astronomical observations allow us to better our understanding of the universe. As we observe smaller and more distance features, we run into the diffraction limit of our observation system. This limit is a function of the wavelength observed and the size of the primary aperture used. We can synthesize a larger primary aperture by implementing interferometry. Long baseline optical interferometry would lead to significant improvements in astronomical imaging resolution. Building optical interferometers with free space baselines becomes significantly more difficult as the baseline increases. A promising alternative is to use optical fiber to connect telescopes. Fiber-based interferometers are much more susceptible to phase noise than their free-space counterparts due to inhomogeneities in the fiber medium. This leads to significant degradation of interferometric signals used for astronomical measurements. We implement phase stabilization techniques used in quantum communications to stabilize an optical interferometer with a 170~km pseudo-baseline. We use a quantum optimal measurement technique with this interferometer to resolve the extent of a source four times smaller than the diffraction limit of the system within 1.5\% error. These results bring the potential for a full-scale on-sky long baseline interferometer significantly closer. A 350~km baseline optical interferometer at 1550~nm would allow us to resolve sub microarcsecond features in the universe.

\end{abstract}

\maketitle

Classical imaging systems are limited by the size of their primary aperture and the wavelength of light used \cite{Rayleigh1879}. This limit is known as the diffraction limit. Interferometry allows us to create a larger synthetic primary aperture for a telescope that allows us to observe with much higher resolution than with a single optic, while still being diffraction limited \cite{Thompson2017}. The first long baseline interferometers were established in the radio regime. Since we are able to record the signals from these telescopes, we can interfere them at a later time. This allows us to construct very long baseline radio interferometers (VLBI) such as the Very Long Baseline Array (VLBA), the Square Kilometre Array (SKA), and the Event Horizon Telescope (EHT), a combination of many independent telescopes and telescope arrays. The EHT was able to resolve features of near 20~microarcseconds and thus produce an image of the black hole in M87 \cite{Akiyama2019First}. A transition to the optical regime would allow us to observe with orders of magnitude higher resolution. This then becomes limited by the fact we can no longer record the signals in real time as they do in radio astronomy. The largest existing optical interferometer is the Center for High Angular Resolution Astronomy (CHARA) array, with a maximum baseline of 331~m \cite{Brummelaar2005}. Each of the telescopes in the array are connected centrally via vacuum tubes to avoid phase noise being accrued by the light. Steps are being taken with the upcoming 7th telescope in the array to integrate optical fiber in order to extend this baseline to up to 1~km \cite{Koehler2024Integrating}. The move towards fiber connected optical interferometers comes with the challenge of more stringent phase stabilization requirements.

Active phase stabilization systems have been significantly more prevalent as we move further towards free-space optical communication (FSOC) and quantum key distribution (QKD). QKD has similar problems to astronomy in which you have a low light signal whose quantum state needs to be preserved over a long distance. Clivati et al. \cite{Clivati2022} and Pittaluga et al. \cite{Pittaluga2021} used a system of off-band stabilization that allows the quantum signal to remain untouched, while sending a separate high powered stabilizing signal down the same optical path (fiber optic cable) so that it experiences the same phase noise. The frequency of this stabilizing signal is similar to that of the quantum signal, thus correcting for the noise on the stabilizing signal corrects for noise on the quantum signal to first order. This stabilizing system is able to match the paths of two arms of an optical interferometer so that interferometry can be performed over hundreds of kilometers. Building an interferometer of this scale would allow for diffraction limited imaging with resolution less than one microarcsecond. Imaging at this resolution allows us to study compositions of exoplanets, star system formation, perform precise geodesy, and perform tests of the General Theory of Relativity through more detailed observations of the regions around black holes such as that in M87 \cite{Huang2026}. Observing faint sources can be difficult with long baseline interferometry due to photon loss in the fiber. This means the measurement should extract as much information as it can from each photon received. Research has gone towards producing quantum estimators that will outperform existing classical estimators \cite{Giovannetti2011, Pearce2017, Huang2021} . We see the technique proposed by Pearce et al. \cite{Pearce2017}, experimentally verified by Howard et al. \cite{Howard2019}. The technique proposed by Huang et al. \cite{Huang2021} has been experimentally verified by Zanforlin et al. \cite{Zanforlin2022}. These works have demonstrated that they are able to outperform classical analysis, but this has not been demonstrated over long baselines. Additionally, advancements in hollow core fibers \cite{Chen2024}, and ZBLAN fiber drawn in microgravity \cite{Starodubov2014} look to have significantly lower loss, which will make the construction of optical VLBI arrays much more practical.

\begin{figure*}[hbt!]
    \centering
    \includegraphics[width=0.97\linewidth]{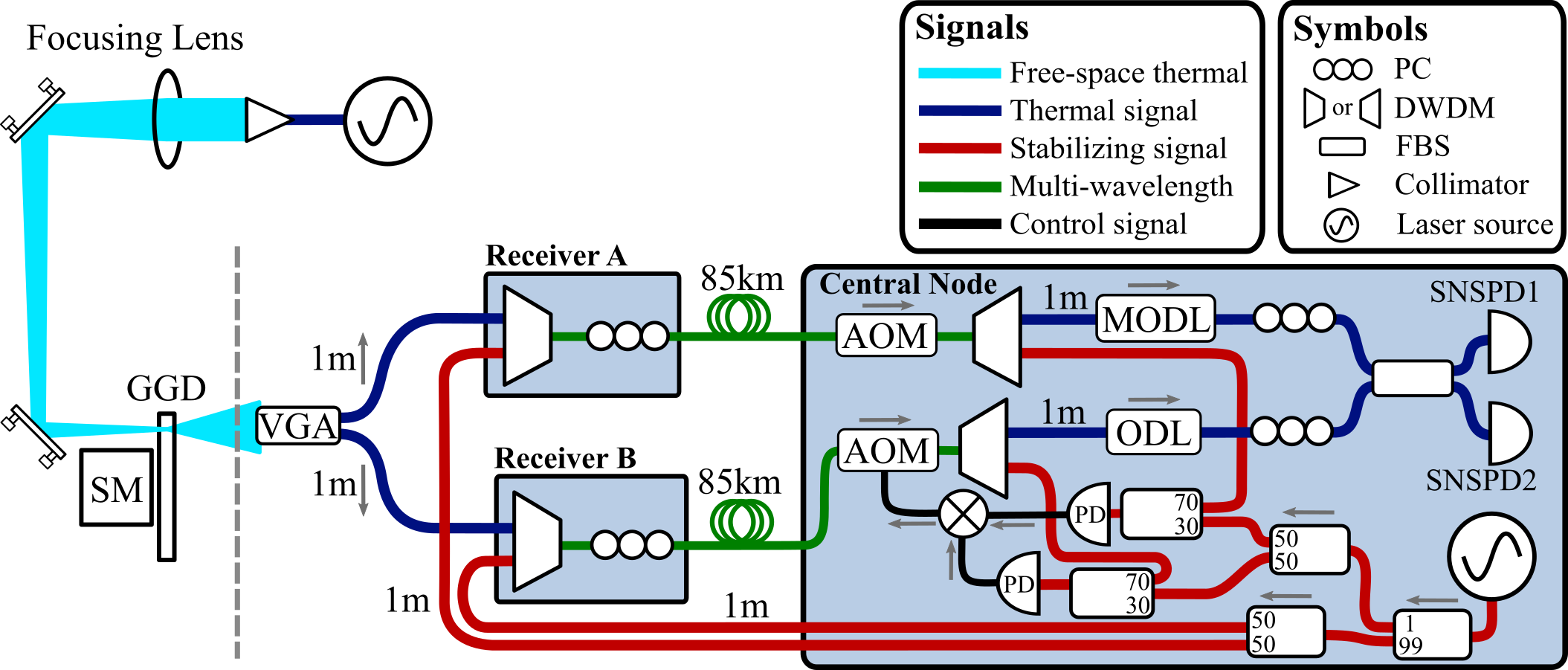}
    \caption{Simplified schematic of the pseudo-thermal source generation and stabilized interferometer. Left: Pseudo-thermal source generation. Light-blue is the thermal source in free space. Right: Optical interferometer. VGA: v-groove array, AOM: acousto-optic modulator, (M)ODL: (motorized) optical delay line, SNSPD: superconducting nanowire single photon detector, PD: photodetector, PC: polarization controller, DWDM: dense wavelength division multiplexer, FBS: fiberized beam splitter, GGD: ground glass disk, SM: stepper motor. The FBS split ratios are listed within the symbol where relevant.}
    \label{fig:system_diagram}
\end{figure*}

Stellar sources are difficult to emulate in a laboratory for testing purposes. Some groups use warm vapor cells \cite{Mika2018}, phase and amplitude noise added to a coherence source \cite{Zanforlin2022}, or use a rotating ground glass disk (GGD) \cite{Howard2019, Liu2021}. This GGD method is important since the source can still be narrowband while inducing spatial incoherence. This is because astronomical sources, which are spectrally broad, will experience significant chromatic dispersion over 85~km of optical fiber. Nevertheless, this effect is linear and should be easily mitigated through the use of dispersion compensating components \cite{Gul2023} in a full scale experimental application of this technique. Our previous work demonstrates that the stabilization system is still sufficient to stabilize a wideband signal \cite{Collier2026}. This work aims to use a narrowband spatially incoherent source to emulate a thermal source, and recover its size with a quantum optimal estimator.

The interferometer starts with two receivers which are two single mode fibers from a v-groove array (VGA). The VGA has four single mode fiber ports separated by 127~\textmu m, thus we can create three different baselines (127~\textmu m, 254~\textmu m, and 381~\textmu m) to take the measurements. We take two of these ports as the inputs to the interferometer. Once two of the ports are selected, the signal from each of them is combined with a stabilization probe signal using a dense wavelength multiplexer (DWDM). Each of these arms are connected to 85~km of optical fiber that travels around Boorloo (Perth, Western Australia). Once these signals return, each arm is connected to an acousto-optic modulator (AOM). These are driven with an 85~MHz tone, one of which is modulated such that it is driven to follow the phase noise of the other arm. The signal out of the AOM is then split by a DWDM so we have isolated quantum and stabilizing signals. Each arm's stabilizing signal is interfered with a local oscillator to measure phase noise on the link. The difference of these is used to stabilize the links relative to each other. The quantum signal from each arm travels through optical delay lines, one of which is motorized so that we can induce a relative phase difference. These signals are then interfered on a beam splitter before being measured by the superconducting nanowire single photon detectors (SNSPDs). This single photon amplitude interferometry allows us to collect phase information from the source. The SNSPD signals are recorded with the Time and Frequency Analyzer tool of a Moku:Pro, which also handles the phase stabilization via the in-built Lock-In Amplifier tool. 

We produce the pseudo-thermal source by a rotating 1500 grit GGD. This disk is fixed to a stepper motor controlled via an Arduino. 
We collimate a coherent laser and send it into a 250~mm focal length lens. This beam is then focused on the GGD. The scatter from this disk is the spatially incoherent source that can be used to emulate a thermal source \cite{Martienssen1964}. Critically this source is still temporally coherent and thus we can avoid chromatic dispersion that would otherwise obscure the interference and thus reduce the measured complex degree of coherence (CDC). The fringe visibility is calculated by taking the magnitude of the CDC, which can then be used to calculate the size of the source. The path lengths of the two arms are still matched in order to simulate a full scale optical interferometer. We found it important to point the VGA slightly off-axis on the GGD, as there was a central coherent region that produced fringe visibility of 100\% while the GGD is rotating, where we would expect degradation. We take measurements to optimize the polarization of each arm without the disk rotating for optical interference. We also balance the power in the receivers while the disk is stationary and then while it is rotating, as starting the stepper motor can lead to slight misalignment in the optical path.

Following \cite{Pearce2017} and \cite{Howard2019} we define $P(n,m)$ as the probability of detecting $n$ photons in SNSPD1, and $m$ photons in SNSPD2 (from Figure \ref{fig:system_diagram}, within a given measurement window. Since we have single photon detectors that cannot detect multi-photon events, we limit the space to $[n,m] = [1,0], [0,1],$ and $[1,1]$. We derive a series of equations that define the probability of detection in each of the states, based on Pearce et al. \cite{Pearce2017}. 

\begin{gather}
\begin{aligned}
    P(1,0)=\alpha_+ + \alpha_-\cos\phi\\
    P(0,1)=\alpha_+ - \alpha_-\cos\phi\\
    P(1,1)=\beta_+ + \beta_-\cos 2\phi,
    \label{eq:quantum}
\end{aligned}\\
\alpha_\pm=[p_{in}(1,0)\pm p_{in}(0,1)],\\
\beta_\pm = \frac{1}{2}[\pm p_{in}(2,0)\pm p_{in}(0,2)+p_{in}(1,1)], \\
p_{in}(n,m)=\frac{z_+^n}{(1+z_+)^{n+1}}\frac{z_-^m}{(1+z_-)^{m+1}},\\
z_{\pm}=\overline{n}(1\mp|\gamma|).
\end{gather}

Where $\phi$ is the phase offset between the two arms of the interferometer, $\overline{n}$ is the mean number of photons, and $\gamma$ is the CDC. Derivations of these equations can be found in Appendix B. From here, we produce the visibilities and compare to the theoretically expected value for the given baselines to produce an estimate of the source size. 

\begin{figure}[hbt!]
    \centering
    \includegraphics[width=0.97\linewidth, trim=0.4cm 0.4cm 0.5cm 0.45cm, clip]{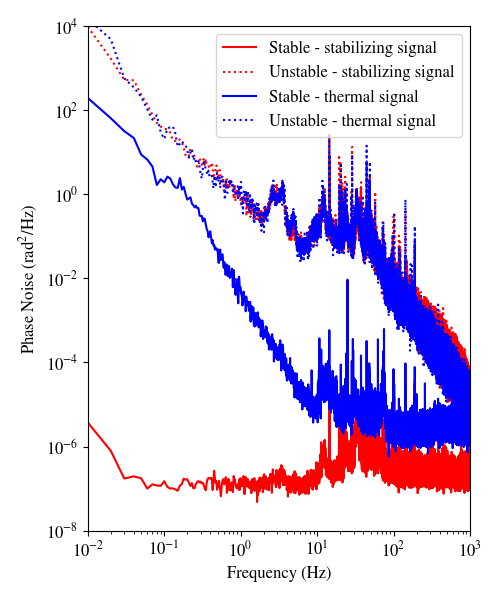}
    \caption{Phase spectral density of interferometer. The red curves are stabilizing signal, and the blue curves are thermal signal. The dashed lines show the response of the system without stabilization, and the solid lines show the response of the system with stabilization. Reproduced from \cite{Collier2026}.}
    \label{fig:psd}
\end{figure}

As demonstrated in figure \ref{fig:psd}, the phase stabilization system is able to reduce a significant amount of the phase noise on the optical link, thus allowing us to recover appropriate visibility values for the interferometer. The kick up in phase noise at low frequencies is due to the offset between the stabilizing signal and the thermal signal. This is required so that the beams can be separated via DWDM. The separation of stabilizing and thermal happens below 10~Hz, we can calculate the expected uncorrelated phase noise $S_{min}(f)$ at frequency $f$ with the equation below \cite{Bertaina2024}.

\begin{equation*}
    S_{min}(f)=\frac{(\lambda_s-\lambda_t)^2}{\lambda_s^2}\frac{lL}{f^2}
    \label{eq:phase_noise}
\end{equation*}

where $l$ is a coefficient of noise in fiber, $L$ is the length of the arms, $\lambda_s$ is the stabilizing laser wavelength, and $\lambda_t$ is the thermal source wavelength.

\begin{figure}[hbt!]
    \centering
    \includegraphics[width=0.97\linewidth, trim=0.55cm 0.55cm 0.5cm 0.45cm, clip]{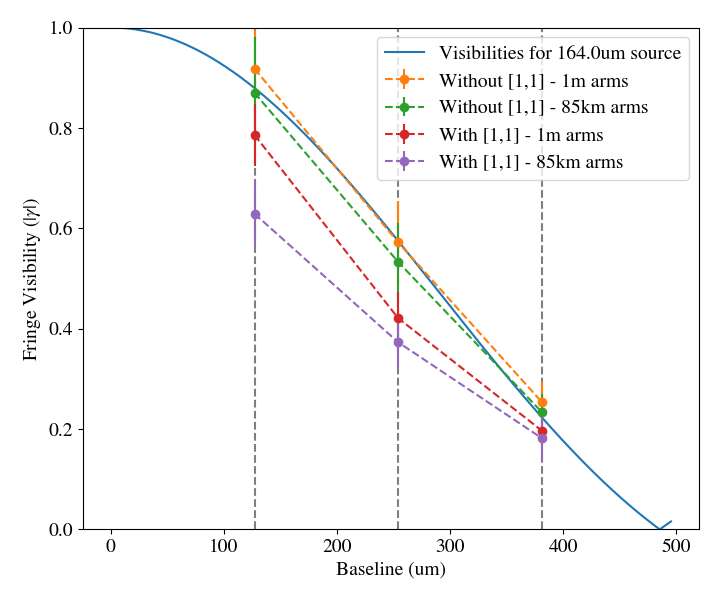}
    \caption{Complex degree of coherence measured at a series of baselines. We have a 164~\textmu m across source 42~mm away from the receivers of the interferometer. Error bars are produced by fitting a series of curves to chunks of the collected dataset, then taking the best fit for each of them, the mean of these is the CDC estimate, the standard deviation is the error. Source size estimate from green curve is 166.47~\textmu m.}
    \label{fig:42mm_visibilities}
\end{figure}

Figure \ref{fig:42mm_visibilities} demonstrates that, with phase stabilization switched on, the introduction of the 85~km of fiber to each arm has minimal impact on the recorded CDC for each of the baselines. Using only the [0,1] and [1,0] counts we recover an estimated source diameter of 166.47~\textmu m, this is within 1.5\% of the actual source size of 164~\textmu m. We find that including the coincidence counts (the [1,1] counts) significantly reduces the CDC. The low number of coincidence counts combined with the high level of noise degraded the performance of the curve fitting of equation \ref{eq:quantum}.


\begin{figure}[hbt!]
    \centering
    \includegraphics[width=0.95\linewidth, trim=0.55cm 0.55cm 0.5cm 0.5cm, clip]{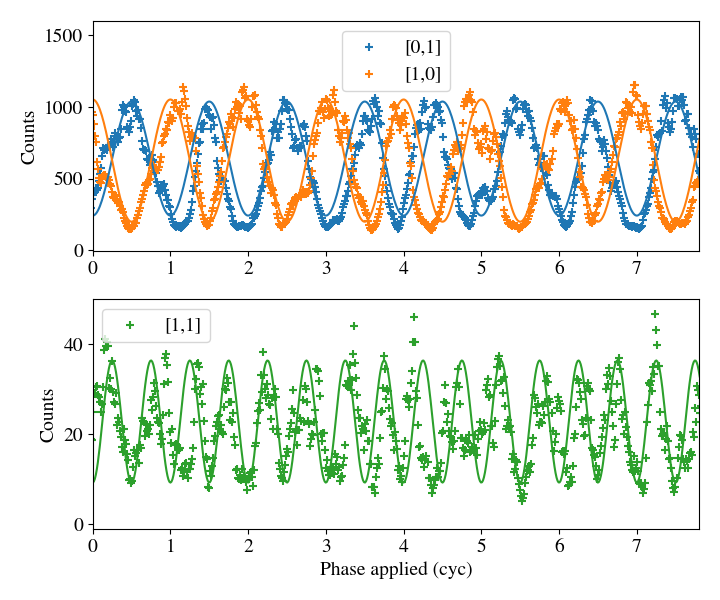}
    \caption{Phase sweep of the stabilized interferometer with 127~\textmu m collecting baseline, and a 164~\textmu m source that is 42~mm away. Orange and blue curves are the first order interference fringes, the green curve is the second order. Solid curves are sinusoids matched to the data.}
    \label{fig:127um_stab}
\end{figure}

\begin{figure}[hbt!]
    \centering
    \includegraphics[width=0.95\linewidth, trim=0.55cm 0.55cm 0.45cm 0.30cm, clip]{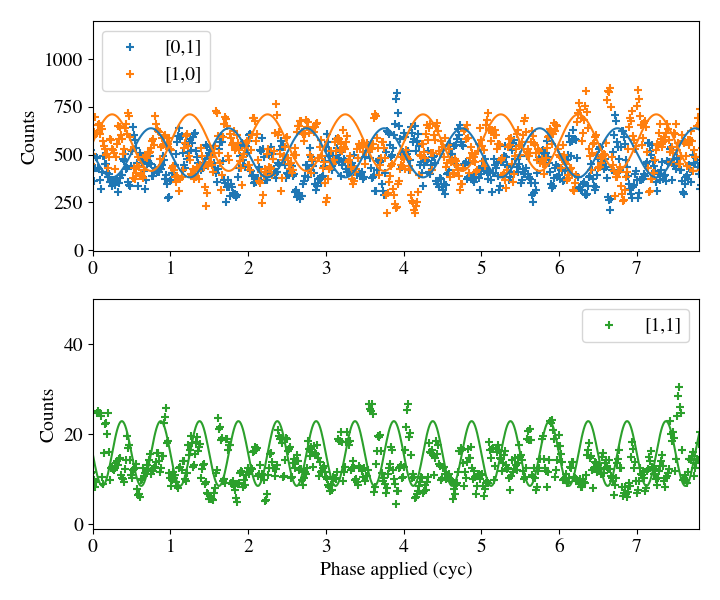}
    \caption{Phase sweep of the interferometer before introducing the stabilization system with a 127~\textmu m collecting baseline, and a 164~\textmu m source that is 42~mm away. Orange and blue curves are the first order interference fringes, the green curve is the second order. Solid curves are sinusoids matched to the data.}
    \label{fig:127um_nostab}
\end{figure}

The times series in figures \ref{fig:127um_stab} and \ref{fig:127um_nostab} demonstrate the need for phase stabilization. Without stabilizing, the noise coupled into the optical fibers is coupled into the quantum signal and the interference is destroyed. Recovering the interference via stabilization lets us estimate the size of the source using the quantum estimator. The source is 3.8 times smaller than the diffraction limit of the system at the shortest baseline, and 1.3 times smaller in the longest. 

In Figure \ref{fig:127um_nostab} we see a large variance in intensity, this is because the light is still temporally coherent, thus even if the path length of the arms are different, the light from them will still interfere. This interference is none the less not useful as it does not correlate with our applied phase, and the visibility is greatly reduced. This is not seen for sources that are wideband \cite{Collier2026}. These variations in intensity are not coupled to the applied phase and thus we are not able to extract any information out of them. 

In conclusion, we use a quantum optimal estimator to measure the size of a pseudo-thermal source with a 170~km pseudo-baseline optical interferometer. Phase stabilization of the interferometer has allowed the interferometric fringes to be recovered when they were previously destroyed by phase noise coupled into the optic fibers. This opens the path for very long baseline astronomical optical interferometers. If this system were to be implemented with the two receivers physically separated by 170~km, the classical diffraction limit would allow us to resolve sources around 2~microarcseconds.

Though correlated [1,1] fringes are recovered in Figure \ref{fig:127um_stab}, they have too much noise coupled in to be functional with the current system, meaning we rely entirely on first order interference. This still allows us to use a quantum optimal estimator to calculate the size of the source. Combined with work from \cite{Collier2026}, we have demonstrated that fringe visibilities can be recovered from a spatially incoherent narrowband source, and can be recovered from a temporally incoherent wideband source. The present results indicate nothing to prevent the construction of an interferometer with a baseline of 330~km which, observing at 1550~nm, would be able to resolve sub-microarcsecond features. Such an instrument would have outstanding applications in the study of exoplanets, star- and planet-forming regions, and the event horizons of super-massive black holes \cite{Huang2026}. The practical extent of the sub-diffraction imaging capability promised by the quantum-optimal measurement method remains to be demonstrated on real-world astronomical targets where the noise resulting from background light and other sources will negatively impact the quantum estimator.


\vspace{0.3cm}\textit{Acknowledgments} - J.J.C and J.S.W are supported by Australian Government Research Training Program Scholarships. D.R.G. is supported by a DECRA Fellowship (DE24010058). This work was supported by the Australian Research Council Centre of Excellence for Engineered Quantum Systems (CE17010009). This material is based upon work supported by the Air Force Office of Scientific Research under award number FA2386-23-1-4081. The authors thank AARNet for the provision of light-level access to their fiber network infrastructure.

\vspace{0.3cm}\textit{Data Availability} - Data underlying the results presented in this paper are not publicly available at this time but may be obtained from the authors upon reasonable request.

\bibliography{references}

\appendix

\newpage
\section{Appendix A: Second order correlations}
\begin{figure}[hbt!]
    \centering
    \includegraphics[width=0.95\linewidth]{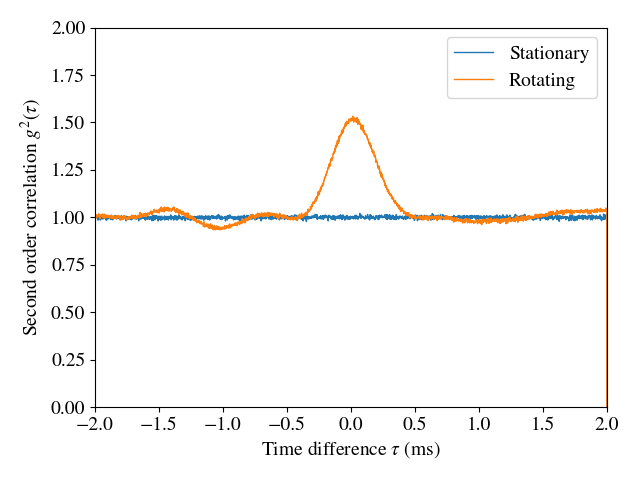}
    \caption{Measurement of second order coherence of the thermals source without and with rotating the GGD.}
    \label{fig:g2}
\end{figure}

Figure \ref{fig:g2} displays the difference in second order correlation between when the ground glass disk is rotating and when it is stationary. The $g^2(\tau)$ of the stationary disk is as expected, a perfectly coherent beam with a poissonian distribution, thus giving is a uniform normalized value of 1. The rotating disk instead has a peak at $\tau\approx0$~ms, demonstrating that it is a super-poissonian source. We see value of $g^2(0)=1.51$, this means we expect thermal behavior in our photon distribution. For a truly thermal source we expect this peak to hit a value of 2. 

\section{Appendix B: Detection probability derivations}

This derivation follows Appendix C of \cite{Pearce2017}; we re-derive the equations for the fringes of interest because the result as written by Pearce et al. includes singular terms that are difficult to systematically remove. Defining creation operators $\hat{a}^\dagger_1,\hat{a}^\dagger_2$ that create photons at receiver A and receiver B of the interferometer respectively and assuming that the mean photon number at each port is equal, the second order correlations between these operators are
\begin{gather*}
    \langle\hat{a}^\dagger_A\hat{a}_A\rangle = \langle\hat{a}^\dagger_B\hat{a}_B\rangle = \overline{n},\\
    \langle\hat{a}^\dagger_B\hat{a}_A\rangle = \overline{n}|\gamma|e^{i\phi},
\end{gather*}
with all other combinations vanishing. The $\phi$ that appears here should be thought of as the difference between the true CDC phase and the applied phase shift $\phi_a$. We define a unitary transformation:
\begin{gather*}
    \hat{b}_i = U(\phi)_{ij}\hat{a}_j,\\
    U(\phi) = \frac{1}{\sqrt{2}}\begin{pmatrix}
    -e^{-i\phi}&e^{-i\phi}\\
    1&1
    \end{pmatrix}
\end{gather*}
The new modes obey the creation/annihilation operator commutation relations and satisfy
\begin{gather*}
\langle\hat{b}^\dagger_A\hat{b}_A\rangle = \overline{n}(1-|\gamma|),\\
\langle\hat{b}^\dagger_B\hat{b}_B\rangle = \overline{n}(1+|\gamma|),\\
    \langle\hat{b}^\dagger_B\hat{b}_A\rangle = 0.
\end{gather*}
These operators are decoupled to second order; we postulate that $\hat{b}_A,\hat{b}_B$ are uncorrelated thermal modes with mean photon numbers $\overline{n}(1\mp|\gamma|)$. As a consequence, the probability of finding n photons in mode $\hat{b}_A$ and m photons in mode $\hat{b}_B$ is
\begin{gather*}
    p_{\text{in}}(n,m) = \frac{z_+^n}{(1+z_+)^{n+1}}\frac{z_-^m}{(1+z_-)^{m+1}},\\
z_{\pm}=\overline{n}(1\mp|\gamma|).
\end{gather*}
The states of interest are those with n photons in one detector and m in the other after the action of the beamsplitter, given by $|n,m\rangle := \frac{1}{2^{n+m}\sqrt{n!m!}}(\hat{a}_B+\hat{a}_A)^n(\hat{a}_B-\hat{a}_A)^m|0\rangle$. Inverting the unitary transformation above lets us write these states in terms of the $b$ modes. Doing this explicitly for the lowest three fringes gives
\begin{gather*}
|1,0\rangle = \frac{1}{2}\left[(1-e^{i\phi})|1,0\rangle_b + (1+e^{i\phi})|0,1\rangle_b\right],\\
|0,1\rangle = \frac{1}{2}\left[(1+e^{i\phi})|1,0\rangle_b + (1-e^{i\phi})|0,1\rangle_b\right],\\
|1,1\rangle = \frac{1}{4}\big[\sqrt{2}(1-e^{2i\phi})|2,0\rangle_b+\sqrt{2}(1-e^{2i\phi})|0,2\rangle_b\\\qquad+2(1+e^{2i\phi})|1,1\rangle_b\big].
\end{gather*}
Using the Born rule in the form $P(n,m) = \sum_{n'm'}|\langle n,m|n',m'\rangle_b|^2 p_{\text{in}}(n',m')$ gives the result quoted above.

\end{document}